# Nanostructure and related mechanical properties of an Al-Mg-Si alloy processed by severe plastic deformation.

Gulnaz Nurislamova[a,b], Xavier Sauvage[a**], Maxim Murashkin[b], Rinat Islamgaliev[b*], Ruslan Valiev[b]

[a] *Groupe de Physique des Matériaux, UMR CNRS 6634, Institute of Material Research, Université de Rouen, 76801 Saint-Etienne du Rouvray, France.*
[b] *Institute for Physics of Advanced Materials, Ufa State Aviation Technical University, K.Marx 12, Ufa 450000, Russia.*

**Abstract**
Microstructural features and mechanical properties of an Al-Mg-Si alloy processed by high-pressure torsion have been investigated using transmission electron microscopy, X-ray diffraction, three-dimensional atom probe, tensile tests and microhardness measurements. It is shown that HPT processing of the Al-Mg-Si alloy leads to a much stronger grain size refinement than of pure aluminium (down to 100 nm). Moreover, massive segregation of alloying elements along grain boundaries is observed. This nanostructure exhibits a yield stress even two times higher than that after a standard T6 heat treatment of the coarse grained alloy.

_______________________
Corresponding authors:

* Rinat Islamgaliev: saturn@mail.rb.ru

**Xavier Sauvage: xavier.sauvage@univ-rouen.fr

-----------------------

**Introduction**

During the last decade, severe plastic deformation (SPD) techniques have been successfully applied to a large number of metallic alloys to achieve ultrafine-grained structures with unique properties [1,2]. Thanks to the refinement of coarse-grained structures down to the nanoscale, high strength and superplastic properties are commonly achieved in light alloys. Among commercial aluminum alloys, the age hardenable 6061 is widely used because it exhibits a good combination of formability, strength, corrosion resistance and weldability. To further improve its mechanical properties, several authors tried to reduce the grain size by SPD using equal channel angular pressing (ECAP) [3-6] and the scaling up of this procedure was also successfully applied [7]. A significant increase of the strength is reported, especially for pre-ECAP solid-solution treatment added by post-ECAP aging treatment [4-6,8]. Moreover, single pass ECAP processing is reported to improve the fatigue properties [9]. Another SPD technique the so-called High Pressure Torsion (HPT) is known to achieve enhanced grain refinement [1,2], causing a further increase of the mechanical properties of the 6061 aluminum alloy. The aim of the present study was to apply the HPT process to reach a maximum strength for the aluminum alloy. Both mechanical properties and microstructures were investigated in order to find the connections presently possible by SPD.

**Experimental Methods**

The alloy investigated in the present study is a hot pressed commercial 6061 aluminum alloy (Mg 0.8-1.2, Si 0.4-0.8, Cu 0.15-0.4, Cr 0.15-0.35, Mn 0.15, Fe 0.7, Zn 0.25, Ti 0.15 (wt.%)), 20 mm in diameter and 150 mm in length. Before HPT the alloy was heat treated at 530°C for 2 hours and then water quenched in ice brine. After this treatment, the grains are equiaxed with a mean size of 100 μm (not shown here). The solution treated billet was cut in two parts. The first part was subjected to the standard T6 ageing treatment (170°C during 12 hours). From the second part of the billet, samples (diameter 20 mm and thickness 0.3 mm) were cut and then HPT processed (5 rotations under a pressure of 6 GPa) [1] at room temperature. The true logarithmic strain the HPT processed samples equal to e ~ 7, the shear strain to $\gamma = 1047$. Tensile tests were performed two weeks after HPT processing and one should note that during this time natural ageing occurred. The mechanical properties of the samples were investigated using micro-hardness measurements and tensile tests. The microhardness measurements were recorded along two perpendicular diameters using an indentation distance of 1.25 mm and a force of 3 N. The variation of the results of the microhardness (Vickers hardness HV) measurements was 1 %. Tensile tests have been performed at room temperature at a strain rate of $10^{-4}$ s$^{-1}$ on a computer-controlled testing machine operating with a constant displacement of the specimen grips. The gauge of the tensile test samples was 1.2 mm in length, 1 mm in width and 0.3 mm in thickness. The standard deviation of the ultimate tensile strength and yield stress did not exceed 5%.

The microstructure was investigated by X-ray diffraction (XRD), transmission electron microscopy (TEM) and atom probe tomography (APT). X-ray diffraction patterns were recorded with a Pan Analytical X"Pert diffractometer using CuKα radiation (50 kV, 40 mA). The internal lattice strains ($<\varepsilon^2>^{1/2}$) and the size of coherent domains ($D_{XRD}$) were determined from the broadening of the lines using the modified Williams-Hall method [10-12]. The method by Williamson-Hall is used in case when X-ray peaks corresponding to reflections of different order from one family of planes are not available or do not possess a shape favorable for representation by a Fourier series [1]. TEM samples were prepared by double jet-electropolishing in a solution of nitric acid-methanol (1:3) at a temperature of -30±2°C using a voltage of 13V. The thin foils were examined in a JEOL 2000FX electron microscope operating at 200 kV. Selected area electron diffraction (SAED) patterns were taken with an aperture of 1 $\mu m^2$. The grain size distribution was determined from dark field images with at least 50 grains. APT specimens were prepared by electropolishing (20°C, 10V, 2% perchloric acid in 2-butoxyethanol). APT investigations were carried out with a CAMECA energy compensated atom probe equipped with an advanced delay line detector [13]. Analyses were performed in UHV conditions at 80 K with a pulse fraction of 19 % and a repetition rate of 2 kHz.

**Results and discussion**

As seen in table 1 the HPT processing leads to a strong increase of the hardness (table 1). This hardness is almost 50% higher after HPT than in the T6 state, while Kim and co-authors reported an increase of only about 25% in a similar alloy processed by ECAP [4]. One could argue that the shear strain during HPT is much higher than during ECAP processing. However these authors also reported that there is a saturation of the hardness after only few ECAP passes (25 % increase of hardness is the maximum achieved at 4 passes). Hardness profiles measured across the sample clearly revealed that the hardness is homogenous across the sample (data not shown here), although the shear strain $\gamma$ is a function of the radius ($\gamma = 2\pi N r / h$, where N is the number of rotation, r the radius and h the sample thickness [1]). This feature is consistent with previous published data of pure aluminum showing that already four HPT rotations (e ~6, $\gamma$ =502, at r=5 mm, h=0.25 mm) lead to a saturation and homogenization of the hardness in the HPT sample for a number of rotation equal or larger than four [14,15]. Since the properties of the material look homogenous in the whole sample after HPT, it makes sense to cut tensile test specimens to record stress-strain curves. A strong increase of the yield stress is exhibited along with a significant drop of the ductility, but elongation to failure nevertheless remains ~ 5.5%. The yield stress of the 6061 Al alloy processed by HPT (660 MPa) is much higher than after the conventional T6 precipitation hardening treatment (275 MPa in this work and 268 MPa in [16]). It is also higher than the yield stress of the same material processed by other SPD techniques like ECAP (386 MPa [4]), ECAP followed by ageing (411 MPa [4]), accumulative roll bonding (363 MPa [17]), ECAP followed by cold rolling (475 MPa [16]) or ECAP used for the consolidation of powders (248 MPa [18]). It is interesting to note that only for two Al alloys a higher yield stress was measured: in Al-7.5%Mg [19] and in Al-4.4%Mg-0.7%Mn-0.15%Cr (Al 5083) [20] (847 MPa, 760 MPa, respectively).

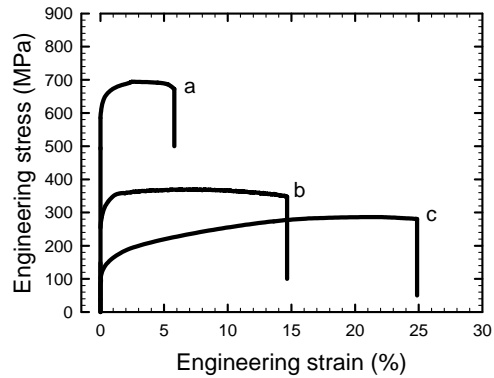

Figure 1: Engineering stress – strain curves of the 6061 Al alloy after high-pressure torsion (a), conventional T6 heat treatment (b) and solution treated and quenched state (c).

These alloys were processed by cryomilling followed by compaction and exhibited a much lower elongation to failure than the present Al alloy 6061 processed by HPT (less than 1.5 % versus 5.5%). This comparison shows that HPT processing leads to very high strength in combination with sufficient ductility. One may note that a T6 aging treatment was applied to the present material processed by HPT but this did not give rise to an additional increase of the yield stress (if you have the data, may be could include it in Fig. 1, what do you think about it?)

Table 1: Microhardness (Hv), yield stress (YS), ultimate tensile stress (UTS) and elongation to failure (El) of the 6061 Al alloy after high-pressure torsion (HPT), conventional treatment (T6) and solution treated and quenched state (ST + WQ).

| Treatment | Hv, MPa | YS, MPa | UTS, MPa | El., % |
|---|---|---|---|---|
| HPT | 1730±18 | 660±21 | 690±28 | 5.5±0.3 |
| T6 | 1175±12 | 276±14 | 365±16 | 14.0±1.0 |
| ST+WQ | 750±8 | 150±7 | 275±10 | 23.0±1.0 |

Figures 2(a) and 2(d) are bright field images of HPT taken from the central region and at 6 mm distance from the sample center. The distance 6 mm from the center is corresponded to position of gage section for specimens subjected to tensile tests and the figure 2 illustrates the structure of these specimens with high strength of 690 MPa. There is no significant difference: grains are equi-axed and the grain size distribution is very similar with a mean grain size of about 100 nm (Fig. 2 (c) and (f)). The existence of Debye-Sherrer rings in SAD patterns (taken from an area of $1\mu m^2$) is an indication that the grain size is in the submicrometer range and HPT deformation leads to the formation of many high-angle grain boundaries (Fig. 2 (b) and (e)). The spreading of spots in the diffraction pattern and the contrast variations inside the grains in the bright field images indicate a significant level of internal stresses and crystal lattice distortions resulting from SPD. Such features are typical of HPT processed materials. They are usually attributed to a high density of dislocations at grain boundaries leading to their high energy and long-range internal stresses [1].

Table 2: Mean coherent domain size ($D_{XRD}$) and internal lattice strain ($<\varepsilon^2>^{1/2}$) of the 6061 Al alloy in the solution treated and quenched state (ST + WQ) and after high-pressure torsion (HPT) in the sample center and at 6 mm distance from the center.

| Treatment | | $D_{XRD}$ (nm) | $<\varepsilon^2>^{1/2}$ (%) |
|---|---|---|---|
| ST+WQ | | - | 0.0100 ±0.0008 |
| HPT | center | 71 ± 4 | 0.1000 ±0.0015 |
| | 6mm from center | 65 ± 3 | 0.1100 ± 0.0020 |

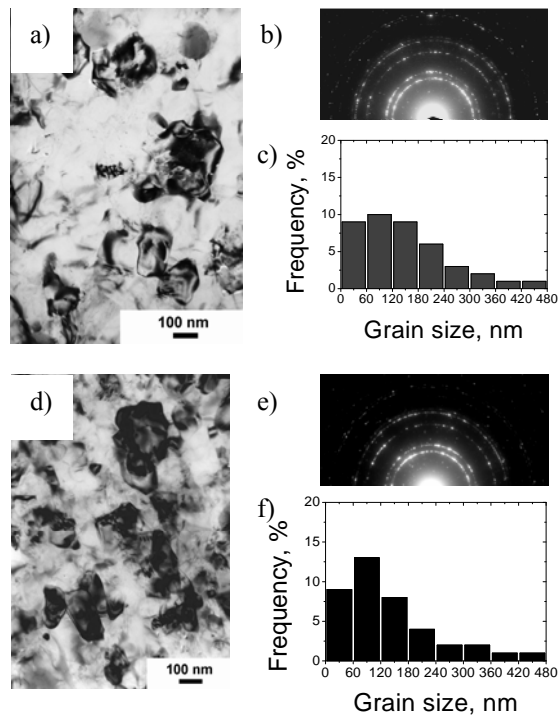

Figure 2: TEM bright field images of the microstructure after HPT, corresponding selected area diffraction patterns (aperture area 1μm$^2$) and grain size distribution in the center of the sample (a, b, c) and at a distance of 6mm from the center (d, e, f).

The size of coherent domains and the mean lattice strains were determined by XRD measurements (Table 2). Although the mean coherent domain size is slightly smaller than the average grain size observed in bright field TEM images, these data are consistent with TEM data and confirm that both the microstructure and the lattice strains are homogenous within the sample processed by HPT.

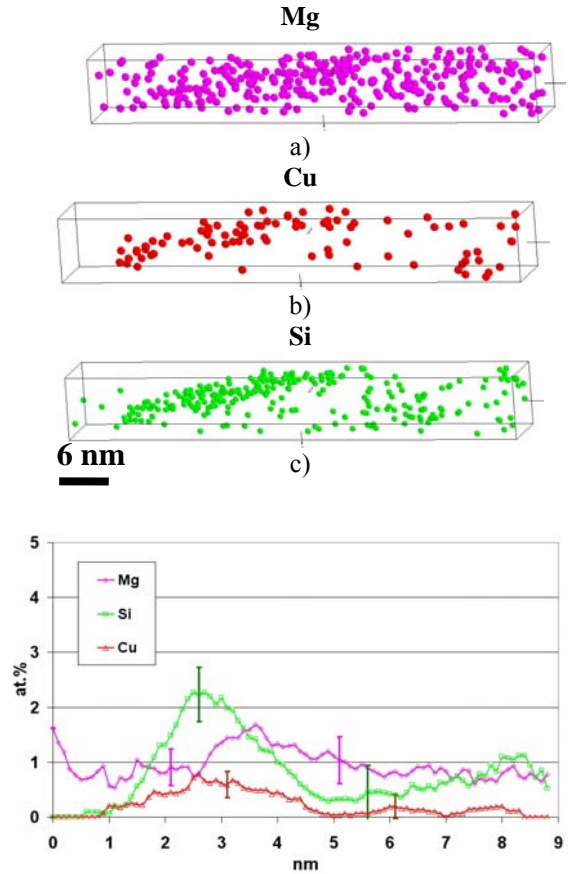

Figure 3: Distribution of Mg atoms (a), Cu atoms (b) and Si atoms (c) in a 3D reconstructed volume analyzed in the 6061 Al alloy processed by HPT (6x6x40 nm$^3$). Segregation of Si, Mg and Cu along a planar defect is exhibited at least by (b) and (c) (why did you change this ?). (d) composition profile computed across the segregation showing that Si concentration is up to 2 at.% and Cu concentration up to 0.7 at.%.

The grain size reduction achieved by HPT in the 6061 Al alloy is much stronger than in the commercially pure Al (about 100 nm in the present alloy versus 800 nm after 8 rotations (e ~ 7, γ =1004, at r=5 mm, h=0.25 mm) in pure Al [14]). A similar small grain size was reported by Islamgaliev and co-authors for two other Al alloys AlMgLiZr [21] and AlZnMgCuZr [22] where a grain size close to 100 nm or even smaller was achieved by HPT. Thus alloying elements seem to play a key role in the microstructure refinement during SPD. Although there is no temperature increase during HPT processing (low strain rate and good thermal contact between the sample and the anvils [1]), alloying elements may decrease the stacking fault energy or may diffuse faster due to SPD induced vacancies [23] or by pipe diffusion [24] and therefore modify the collective behaviour of dislocations and the formation of new grain boundaries.

For comparison the mean grain size 0.3-0.4 μm was observed in the structure of the 6061 alloy processed by ECAP [4,5,16,25] at logarithmic shear strain equal to 12. Microstructure of the 6061 alloy after 8 cycles of rolling of another technique, a namely ARB [17], was characterized by formation of elongated grains of 500 nm in length and 150 nm in width. At the same time as mentioned above the HPT samples have demonstrated the equiaxed grain structure with a mean grain less than 100 nm.

The distribution of Cu, Mg and Si that are in a supersaturated solid solution prior to HPT (other alloying elements are mostly located in dispersoids) was investigated by three-dimensional atom probe. These studies clearly showed that the amounts of Mg and Si in solid solution after HPT (0.8±0.1 at.% and 0.6±0.1 at.%, respectively) are slightly lower than in the solution treated state (1.0±0.1 at.% ; respectively 0.7±0.1 at.%). This is the result of segregation induced by SPD as shown in the analyzed volume displayed in Fig. 3. This is a planar segregation and for more clarity the volume was orientated so that the segregation plane is perpendicular to the figure plane. It is suggested that the HPT process induces a segregation along dislocation cell or grain boundaries. Similar planar segregations were recently reported also for carbon atoms in a pearlitic steel processed by HPT [26].

Using published data on 6061 alloys, one can estimate the Hall-Petch hardening of the 6061 alloy processed by HPT ($\sigma_y = \sigma_o + kd^{-1/2}$, where $\sigma_y$ is the yield stress, $\sigma_o$ and k are two constants and d is the grain size). In the solution treated state (respectively in the ECAP state), the grain size is 100 μm (respectively 350 nm) and the yield stress is close to 150 MPa (close respectively to 400 MPa) [4]. These experimental data are the basis to the following estimates: $\sigma_o$=124 MPa and k=166 MPa*μm$^{1/2}$. Then, applying the Hall-Petch law with a grain size of 100 nm (the present case), the yield stress should be about 649 MPa. This is close to the experimental value (660 MPa), but it is somewhat higher than the estimate and this difference could be attributed to segregations or dislocations at grain boundaries that were observed after HPT processing [25]. Further investigations are planned to clarify this point together with the ageing behavior.

**Conclusions**

The 6061 aluminum alloy processed by HPT exhibits an ultrafine-grained structure with a mean grain size of 100 nm. This microstructure is quite homogeneous within the whole sample. Moreover SPD induces segregations of Mg, Si and Cu along planar defects (supposed to be new grain boundaries), which was clearly revealed by APT analyses. This unique microstructure gives rise to a yield stress twice higher than after a standart T6 heat treatment of the coarse grained alloy. After HPT processing of the 6061 Al alloy a ductility of 5.5% is preserved, what is important for any structural application.

**Acknowledgement**
The authors would like to thank Dr. A.R. Kilmametov for performing X-ray diffraction analyses.